\documentclass[12pt]{elsart}

\usepackage{amssymb}
\usepackage{graphics}
\usepackage{graphicx}
\newcommand{\xx}{\mbox{\boldmath$x$}}

\newcommand{\vv}{\mbox{\boldmath$v$}}

\newcommand{\unitm}{m_{\scriptscriptstyle 0}}
\newcommand{\unitl}{l_{\scriptscriptstyle 0}}
\newcommand{\unitv}{v_{\scriptscriptstyle 0}}
\newcommand{\bftau}{\mbox{\boldmath$\tau$}}
\newcommand{\Tphys}{T_{\rm phys}}
\newcommand{\Cv}{C_{\rm\scriptscriptstyle V}}

\begin{document}

\begin{frontmatter}


\title{\large Gravothermal Catastrophe and Tsallis' Generalized 
Entropy of Self-Gravitating Systems II. Thermodynamic properties  
of stellar polytrope
}
\author[taruya]{Atsushi Taruya}
\address[taruya]{Research Center for the Early Universe(RESCEU), 
School of Science, University of Tokyo, Tokyo 113-0033, Japan}
\ead{ataruya@utap.phys.s.u-tokyo.ac.jp}


\author[sakagami]{Masa-aki Sakagami}
\address[sakagami]{Department of Fundamental Sciences, FIHS, 
Kyoto University, Kyoto 606-8501, Japan}
\ead{sakagami@phys.h.kyoto-u.ac.jp}
\begin{abstract}
In this paper, we continue to investigate the thermodynamic properties of 
stellar self-gravitating system arising from the Tsallis generalized entropy. 
In particular, physical interpretation of the thermodynamic instability,  
as has been revealed by previous paper(Taruya \& Sakagami, 
Physica A 307 (2002) 185), is discussed in detail based 
on the framework of non-extensive thermostatistics. Examining the 
Clausius relation in a quasi-static experiment, we obtain the 
standard result of thermodynamic relation that the 
physical temperature of the equilibrium non-extensive system is 
identified with the inverse of the Lagrange multiplier,  $\Tphys=1/\beta$. 
Using this relation, the specific heat of total system is 
computed, and confirm the common feature of self-gravitating system 
that the presence of negative specific heat leads to the thermodynamic 
instability. In addition to the gravothermal instability discovered 
previously, the specific heat shows the curious divergent 
behavior at the polytrope index $n>3$, suggesting another type of 
thermodynamic instability in the case of the system surrounded by the 
thermal bath. Evaluating the second variation of free energy, we check 
the condition for onset of 
this instability and find that the zero-eigenvalue problem of the 
second variation of free energy exactly recovers the marginal 
stability condition indicated from the specific heat. Thus, the stellar 
polytropic system is consistently characterized by the non-extensive 
thermostatistics as a plausible thermal equilibrium state. We also 
clarify the non-trivial scaling behavior appeared in specific heat
and address the origin of non-extensive nature in stellar 
polytrope. 
\end{abstract}
\end{frontmatter}
%
%
%
%
%
%
%
\section{Introduction}
\label{sec: intro}
%
%
%
%
%
Due to its complexity and peculiarity, 
stellar self-gravitating system has long attracted much attention in 
the subject of astronomy and astrophysics, and even statistical physics. 
For an isolated stellar system, the dynamical equilibrium is rapidly 
attained after a few crossing time and the thermodynamic 
description provides useful information in characterizing the late-time 
behavior of this system. Even in this simplest situation, however, 
the equilibrium state 
of self-gravitating system shows various interesting phenomena, 
which may offer an opportunity to recast the framework of the 
thermodynamics and/or statistical mechanics.

In earlier paper, applying the Tsallis' generalized entropy\cite{T1988}, 
we have studied the thermodynamic instability of self-gravitating 
systems\cite{TS2002}. The self-gravitating stellar system confined in a 
spherical cavity of radius, $r_e$, exhibits an instability, so-called 
{\it gravothermal catastrophe}, 
which has been widely accepted as a fundamental physical process and 
plays an important role for the long-term evolution of globular clusters
\cite{BT1987,EPI1987,MH1997}. 
The presence of this instability has been long known since the 
pioneer work by Antonov\cite{Antonov1962} and 
Lynden-Bell \& Wood\cite{LW1968}. Historically, 
the gravothermal catastrophe has been studied on the basis of the maximum 
entropy principle for the phase-space distribution function, with a 
particular attention to the Boltzmann-Gibbs entropy
\cite{Padmanabhan1989,Padmanabhan1990}.

In contrast to  previous work, we have applied the 
Tsallis-type generalized entropy to seek 
the equilibrium criteria for the first time (for comprehensive review of 
Tsallis formalism and its application to other field of physics, see 
\cite{T1999,AO2001}). Then, 
the distribution function of Vlassov-Poisson system can be reduced 
to a stellar polytropic system\cite{PP1993,PP1999}. Evaluating the 
second variation of entropy around the 
equilibrium state and solving the zero-eigenvalue problem, 
the criterion for the onset of gravothermal instability is obtained. 
The main results of our previous analysis are summarized as follows:

\vspace*{0.5cm}

\noindent
{\bf(i)} Local entropy extremum ceases to exist in cases with polytrope 
index $n>5$ for sufficiently larger radius of the wall, 
$r_e> \lambda_{\rm crit}\,GM^2/(-E)$, and for highly density contrast, 
$\rho_c/\rho_e> D_{\rm crit}$, where $M$ and $E$ denote the total mass 
and energy of the system, $\rho_c$ and $\rho_e$ mean the density at 
center and edge, respectively. 
\\
{\bf(ii)} The critical values $\lambda_{\rm crit}$ and $D_{\rm crit}$ depend 
on the polytrope index, both of which respectively approach 
$0.335$ and $709$ in the limit of $n\to\infty$, 
consistent with the well-known result adopting the 
Boltzmann-Gibbs entropy.  
\\
{\bf(iii)} The stability/instability criterion obtained from 
the second variation of Tsallis entropy exactly matches with the result 
from standard turning-point analysis. 

\vspace*{0.5cm}

While the successful results suggest that 
non-extensive generalization of thermodynamics will offer 
various astrophysical applications involving long-range nature of 
self-gravitating systems, there still remain some important issues 
concerning the physical interpretation of thermodynamic instability.

Heuristically, the gravothermal instability is 
explained by the presence of negative specific heat as follows. 
In a fully relaxed gravitating system with sufficiently larger radius, 
negative specific heat arises at the inner part of the system  
and we have $C_{\rm V,inner}<0$, 
while the specific heat at the outer part remains positive, 
$C_{\rm V,outer}>0$, since 
one can safely neglect the effect of self-gravity. 
In this situation, if a tiny heat flow is momentarily supplied from inner 
to outer part,  both the inner and the outer parts get hotter after the 
hydrostatic readjustment. Now imagine the case, 
$C_{\rm V, outer}>|C_{\rm V, inner}|$.  The outer part has so much thermal 
inertia that it cannot heat up as fast as the inner part, and thereby the 
temperature difference between inner and outer parts increases. As a 
consequence, the heat flow never stops, leading to a catastrophe 
temperature growth.

While the above thought experiment is naive in a sense 
that we artificially divide the system into the inner and the outer part,  
the argument turns out to capture an essence of the thermodynamic 
instability in cases with the Boltzmann-Gibbs entropy.  
Evaluating the specific heat explicitly, Lynden-Bell and Wood\cite{LW1968} 
showed that the specific heat of the total system should 
be greater than zero at the onset of instability, although the central part 
of this system still has the negative specific heat. Therefore, one can 
naively expect that the self-gravitating system generally exhibits the 
thermodynamic instability associated with the negative specific heat and 
this could even hold in the system characterized by the non-extensive entropy.

To address this issue, 
however, we should remember the following two remarks that have been never 
clarified. First note that there exists a subtle point concerning 
the concept of temperature in the non-extensive thermodynamics. 
Framework of the non-extensive formalism is formally 
constructed keeping the standard result of thermodynamic relations
\cite{CT1991,PP1997,TMP1998}, 
however, the physical temperature, $\Tphys$, might not be simply related 
to the usual one, i.e, the inverse of Lagrange multiplier,   
as has been criticized recently\cite{MNPP2000,AMPP2001}. 
This point is in particular important in evaluating the specific heat.

Second, as has been mentioned by the pioneer work of Lynden-Bell \& 
Wood\cite{LW1968}, self-gravitating system shows various types of 
thermodynamic instability. While our early study deals with the 
stellar system confined within an adiabatic wall, one may replace the 
adiabatic wall with the thermally conducting wall surrounded by 
a heat bath. In this situation, assuming the Boltzmann-Gibbs entropy, 
Lynden-Bell \& Wood showed that no equilibrium state exists for 
sufficiently low temperature and high-density contrast. Note that 
even in this case, the presence of negative specific heat plays an 
essential role for the appearance of instability.

Keeping the above remarks in mind, in this paper, 
we focus on the thermodynamic property of self-gravitating systems 
characterized by Tsallis' generalized entropy.  
For this purpose, we first investigate the thermodynamic temperature of 
the self-gravitating system from the Clausius relation. 
To clarify the physical interpretation of thermodynamic instability,  
the specific heat is computed and a role of negative 
specific is discussed in detail. Then we turn to focus on the 
thermodynamic instability 
in a system surrounded by the heat bath. The stability/instability 
criterion is derived from the second variation of free energy and 
a geometrical construction of marginal stability condition is discussed. 

While the problem considered here includes some general issues that 
are commonly faced with the application of the non-extensive 
thermostatistics, we will tackle this problem based on the {\it old} 
Tsallis formalism using the standard statistical average, 
which is currently un-common 
(e.g., \cite{MNPP2000,AMPP2001}\cite{TMP1998}). The reason why 
we do not adopt the {\it standard} Tsallis formalism using the normalized 
$q$-expectation values is twofold. As has been mentioned in previous paper, 
a naive application of the new formalism apparently shows a problematic 
difficulty in our case of the maximum entropy principle 
(Sec.\ref{sec: polytrope}), while no such difficulties arise when we apply 
the earlier formalism (see ref.\cite{TS2002} in details). Another 
reason is that 
while the new formalism has been deliberately constructed so as to eliminate 
the undesirable divergences in some physical systems \cite{TMP1998}
especially with fractal nature, 
no serious divergences have appeared in our case. 
Precisely speaking, the  physical quantities, e.g. mass and energy,
may have divergence for some equilibrium configuration of 
the self-gravitating system.
In order to remedy this divergence, we confine the system within 
a spherical wall, which is a standard prescription 
in studies for the self-gravitating system \cite{Antonov1962,LW1968}. 
Since even the old Tsallis formalism preserves a consistent framework 
that recovers the usual thermodynamic structure, 
from a more general view of the non-extensive thermostatistics,  
we expect that the present analysis still provides a valuable insight to the 
thermodynamic stability of stellar self-gravitating system.   
Of course, the analysis using normalized $q$-value must play an 
important role in the Tsallis' non-extensive framework 
and we plan to extend our analysis to the one with the new 
formulation near future. In this sense, present work can be 
regarded as a preliminary analysis toward the next step. 
This point will be discussed in the last part 
of this paper, together with some implications.

This paper is organized as follows. in section \ref{sec: polytrope},  
we recast the problem that finds the most probable state of equilibrium 
stellar distribution adopting the Tsallis entropy. 
The main part of this paper is section \ref{sec: non-extensive}, in which 
the thermodynamic properties of stellar polytrope 
are investigated in detail. After identification of the thermodynamic 
temperature, the explicit expression for specific heat is presented and 
the marginally stability condition for the thermodynamic instability 
is investigated in both the adiabatic and the isothermal cases. 
In section \ref{sec: free-energy}, thermodynamic instability in a system 
surrounded by a thermal bath is re-considered by means of the free energy 
and the marginal stability condition is re-derived from the second 
variation of free energy. Furthermore, following the preceding results, 
the origin of the non-extensive nature in stellar 
polytropic system is discussed in section \ref{sec: discussion}. 
Finally, section \ref{sec: summary} is devoted to the summary and discussions.
%
%
%
%
%
%
%
\section{Stellar polytrope as an extremum state of 
Tsallis entropy} 
\label{sec: polytrope}
%
%
%
%
%
%
%
%
%
%
%
%
In this section, we recast the problem finding the most probable 
state of equilibrium stellar system, based on the maximum entropy principle. 
In our previous study, the entropy for the phase-space 
distribution function has been introduced without recourse to the 
correct dimensions. Although this does not alter the stability/instability  
criterion for the stellar equilibrium state, for the sake of the 
completeness and the later analysis, we repeat the same calculation as 
shown in ref.\cite{TS2002}, taking fully account of the correct dimensions.

Suppose a system containing $N$ particles which are confined within 
a hard sphere of radius $r_e$. For simplicity, each particle is 
assumed to have the same mass $\unitm$ and interacts via Newton gravity. 
The problem considered here is to find an equilibrium state 
in an adiabatic treatment. That is, we investigate the equilibrium 
particle distribution in which the particles elastically bounce from the wall, 
keeping the energy $E$ and the total mass $M(=N\unitm)$ constant.

For present purpose, it is better to employ the mean-field treatment 
that the correlation between particles is smeared out and the 
system can be fully characterized by the one-particle distribution function, 
$f(\xx,\vv)$, defined in six-dimensional phase-space $(\xx,\vv)$ 
\cite{TS2002,BT1987}\cite{Antonov1962,LW1968,Padmanabhan1989,Padmanabhan1990}. 
Let us denote the phase-space element as 
$h^3(=l_0^3v_0^3)$ with unit length $\unitl$ and unit velocity $\unitv$. 
Since the distribution function $f(\xx,\vv)$ counts the number of particles 
in a unit cell of phase-space, using the standard definition of the 
statistical average, the energy and the total mass are 
respectively expressed as follows:
\begin{eqnarray}
  \label{eq: energy}
  E\,&=&\, K+U \,\equiv \,\,\unitm
  \int\, \left\{\frac{1}{2}\, v^{2}
  +\,\frac{1}{2}\,\Phi(\xx)\right\}\,f(\xx,\vv)\,\,d^6 \bftau,
\\
\nonumber\\
  \label{eq: mass}
  M\,&=&\,\unitm\,N\, \equiv\,\unitm \int\, f(\xx,\vv)\,\,d^6\bftau \,\,, 
\end{eqnarray}
with the quantity $\Phi$ being the gravitational potential: 
\begin{equation}
  \label{eq: potential}
  \Phi(\xx)=-G\,\unitm\,\int\,\,\frac{f(\xx',\vv')}{|\xx-\xx'|}\,\,
  d^6\bftau'.
\end{equation}
In the above expressions, the dimensionless integral measure $d^6\bftau$ 
is introduced: 
\begin{equation}
  \label{eq: phase_measure}
    d^{6}\bftau \equiv \frac{d^{3}\xx\,d^{3}\vv}{h^3}\,\,\,\,\,\,
    ;\,\, h = l_0\,v_0. 
\end{equation}

Owing to the maximum entropy principle, we explore 
the most probable state maximizing the entropy. The entropy quoted here 
is a quantity defined in the phase-space and it counts the number of 
possible particle state. We are specifically concerned with the equilibrium 
state for the Tsallis entropy \cite{T1988}: 
\begin{equation}
\label{eq: S_q}
  S_{q}=-\frac{N}{q-1}\int\, \left[\left(\frac{f}{N}\right)^{q}
    -\left(\frac{f}{N}\right)\right]\,\,d^6\bftau.  
\end{equation}
Maximizing the entropy $S_q$ under the constraints reduces to the 
following mathematical problem using Lagrange multipliers $\alpha$ and 
$\beta$: 
\begin{eqnarray}
\delta S_q-\alpha\,\delta M-\beta\,\delta E=0,
\label{eq: 1st_variation}
\end{eqnarray}
which leads to \cite{TS2002,PP1993,PP1999}:
\begin{equation}
  \label{eq: df_1}
  f(\xx,\vv)\,\, =\,\,A\,\,
\left[\Phi_0-\Phi(\xx)-\frac{1}{2}v^2\right]^{1/(q-1)}, 
\end{equation}
where the constants $A$ and $\Phi_0$ are respectively given by
\begin{equation}
 A=N\,\left\{\left(\frac{q-1}{q}\right)\unitm\beta\right\}^{1/(q-1)}, 
\,\,\,\,\,\,\,\,\,\,\,\,
\Phi_0= \frac{1-(q-1)\unitm\,\alpha}{(q-1)\unitm\,\beta}.
\label{eq: A_Phi0_constants}
\end{equation}

The one-particle distribution function (\ref{eq: df_1}) is often called 
{\it stellar polytrope}, which satisfies the polytropic equation of state 
\cite{BT1987}\cite{PP1993}. 
The density profile $\rho(r)$ and the isotropic pressure $P(r)$ at the 
radius $r=|\xx|$ are respectively given by 
\begin{eqnarray}
 \rho(r)\, &\equiv& \,\unitm\,\int\,f(\xx,\vv)\,\,\frac{d^3\vv}{h^3} 
\nonumber\\
&=& \,\,4\sqrt{2}\pi\,\,B\left(\frac{3}{2},\,\frac{q}{q-1}\right)
 \,\,\frac{\unitm\,A}{h^3}\,
   \left\{\Phi_0-\Phi(r)\right\}^{1/(q-1)+3/2},
   \label{eq: def_rho}
\end{eqnarray}
and
\begin{eqnarray}
 P(r)\,&\equiv &\,\unitm\,\int\,\frac{1}{3}\,v^2\,f(\xx,\vv)\,\,\frac{d^3\vv}{h^3} 
\nonumber\\
& =&\,\left(\frac{1}{q-1}+\frac{5}{2}\right)^{-1}\, \rho(r)\,
\left\{\Phi_0-\Phi(r)\right\},
   \label{eq: def_pressure}
\end{eqnarray}
with $B(a,b)$ being the $\beta$ function. 
Thus, these two equations lead to the relation 
\begin{equation}
 P(r) =K_n\, \rho^{1+1/n}(r), 
\label{eq: Eq. of state}
\end{equation}
with the polytrope index given by
\begin{equation}
 n = \frac{1}{q-1}+\frac{3}{2}.    
\label{eq: n_q}
\end{equation}
In equation (\ref{eq: Eq. of state}), the dimensional constant $K_n$ is 
introduced:  
\begin{equation}
K_n \equiv \frac{1}{n+1}
\left\{4\sqrt{2}\pi B\left(\frac{3}{2},n-\frac{1}{2}\right) 
\,\frac{\unitm\,A}{h^3} \right\}^{-1/n}. 
\label{eq: def_Kn}
\end{equation}
Note that the above quantity is equivalent to the variable $(n-3/2)T/(n+1)$ 
defined in ref.\cite{TS2002}.

Once provided the distribution function, the equilibrium 
configuration can be completely specified by solving the Poisson equation.   
Hereafter, we specifically restrict our attention to the spherically 
symmetric configuration for $q>1$(or $n>3/2$), in which 
the dynamically stable state is safely attainable and 
the thermodynamic arguments turn out to capture the physical 
relevance \cite{BT1987}.

From the gravitational potential (\ref{eq: potential}), it reads 
\begin{equation}
\label{eq: poisson_eq}  
 \frac{1}{r^2}\frac{d}{dr}\left(r^2\frac{d\Phi(r)}{dr}\right)=
  4\pi G \rho(r).
\end{equation}
Combining (\ref{eq: poisson_eq}) with (\ref{eq: def_rho}), we 
obtain the ordinary differential equation for $\Phi$. 
Equivalently, a set of equations which represent 
the hydrostatic equilibrium are derived using 
(\ref{eq: def_rho}), (\ref{eq: def_pressure}) and (\ref{eq: poisson_eq}):  
\begin{eqnarray}
& \frac{dP(r)}{dr}\,=&\,\,-\frac{Gm(r)}{r^2}\,\rho(r), 
\label{eq: hydro_1}
\\
& \frac{dm(r)}{dr}\,=&\,\,4\pi\rho(r)\,r^2.  
\label{eq: hydro_2}
\end{eqnarray}
The quantity $m(r)$ denotes the mass evaluated at the radius $r$
inside the wall. We then introduce the dimensionless quantities: 
\begin{equation}
\label{eq: dimensionless}
 \rho=\rho_c\,\left[\theta(\xi)\right]^n,\,\,\,\,\,\,
r=\left\{\frac{(n+1)P_c}{4\pi G\rho_c^2}\right\}^{1/2}\,\xi, 
\end{equation}
which yields the following ordinary differential equation: 
\begin{equation}
 \theta''+\frac{2}{\xi}\theta'+\theta^n=0,
\label{eq: Lane-emden_eq}
\end{equation}
where prime denotes the derivative with respect to $\xi$. 
The quantities $\rho_c$ and $P_c$ in (\ref{eq: dimensionless}) 
are the density and the pressure at $r=0$, respectively. 
To obtain the physically relevant solution of (\ref{eq: Lane-emden_eq}), 
we put the following boundary condition:
\begin{equation}
 \theta(0)=1, \,\,\,\,\,\,\,\theta'(0)=0.    
\label{eq: boundary}
\end{equation}
A family of solutions satisfying (\ref{eq: boundary}) is referred to 
as the {\it Emden solution}, which is well-known in the subject of 
stellar structure (e.g., see Chap.IV of ref.\cite{Chandra1939}).

\begin{figure}[h]
  \begin{center}
    \includegraphics*[width=13cm]{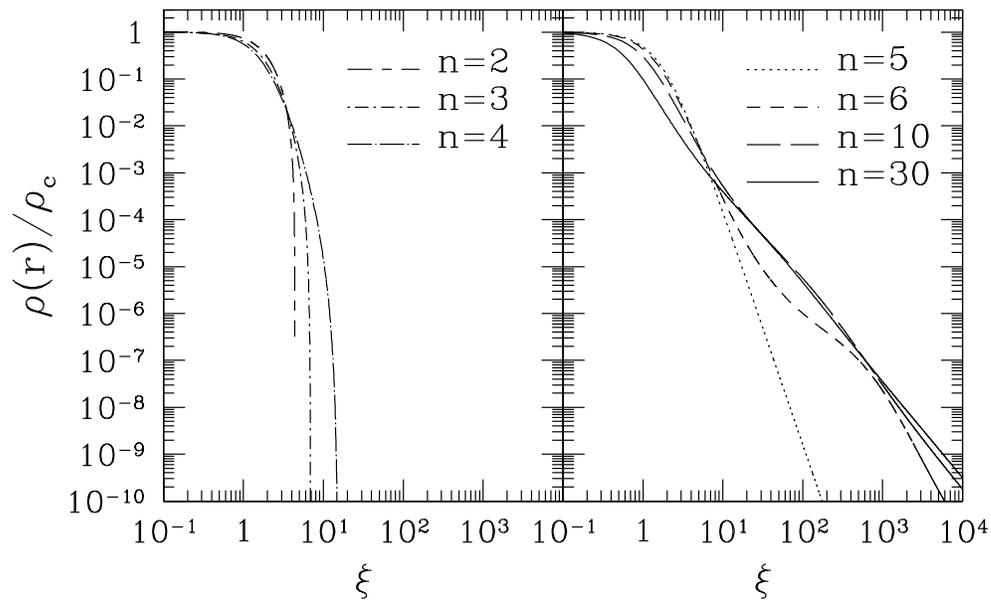}
  \end{center}
    \caption{Density profiles of stellar polytrope for $n<5$
        ({\it left}) and $n\geq5$({\it right}). }
    \label{fig: profile}
\end{figure}

Figure \ref{fig: profile} shows the numerical solution of equation 
(\ref{eq: Lane-emden_eq}) for various polytrope indices, where 
the density profile, $\rho(r)/\rho_c$ is plotted 
as a function of dimensionless radius, $\xi$.  
Clearly, profiles with index $n<5$ rapidly fall off and they 
abruptly terminate at finite radius({\it left-panel}), while 
the $n\geq5$ cases infinitely continue to extend over the outer 
radius({\it right-panel}). As already mentioned in previous paper, 
characteristic feature seen in figure \ref{fig: profile} plays 
an essential role for the thermodynamic instability associated with 
negative specific heat.

For later analysis, it is convenient to introduce the following 
set of variables, referred to as homology invariants 
\cite{Chandra1939,KW1990}: 
\begin{eqnarray}
 u &\equiv& \frac{d\ln m(r)}{d\ln r}=
\frac{4\pi r^3\rho(r)}{m(r)}=-\frac{\xi\theta^n}{\theta'},
\label{eq: def_u}
\\
\nonumber\\
 v &\equiv&  - \frac{d\ln P(r)}{d\ln r}=
\frac{\rho(r)}{P(r)}\,\,\frac{Gm(r)}{r}
=-(n+1)\frac{\xi\theta'}{\theta},  
\label{eq: def_v}
\end{eqnarray}
which reduce the degree of equation (\ref{eq: Lane-emden_eq}) 
from two to one. The derivative of these variables with 
respect to $\xi$ becomes
\begin{equation}
\frac{du}{d\xi} = \left(3-u-\frac{n}{n+1}\,v \right)\,\frac{u}{\xi},
~~~~~~~~
\frac{dv}{d\xi} = \left(-1+u+\frac{1}{n+1}\,v \right)\,\frac{v}{\xi}.
\label{eq: d(u,v)/dxi}
\end{equation}
Equations (\ref{eq: Lane-emden_eq}) can thus be re-written with 
\begin{equation}
 \label{eq: uv_eqn}
  \frac{u}{v}\,\frac{dv}{du}=\frac{(n+1)(u-1)+v}{(n+1)(3-u)-nv}.   
\end{equation}
The corresponding boundary condition to (\ref{eq: boundary}) 
becomes  $(u,v)=(3,0)$. Using these variables, the basic thermodynamic 
quantities such as the energy and the entropy are evaluated and the 
results are summed up in Appendix A, which are subsequently used  
in section \ref{sec: non-extensive}. 
%
%
%
%
%
%
%
%
\section{Thermodynamic properties of stellar polytrope} 
\label{sec: non-extensive}
%
%
%
In this section, we address our main issue, i.e, the physical 
interpretation of gravothermal instability in stellar polytropes, 
based on the framework of non-extensive thermodynamics. 
In section \ref{subsec: temperature}, 
we first discuss the thermodynamic temperature of stellar polytrope 
calculating both the heat and the entropy changes 
in a quasi-static treatment. Then we evaluate the specific heat in section 
\ref{subsec: specific heat}. The connection between 
the absence of extremum entropy state and the presence 
of negative specific heat is discussed in detail. Further,  
we argue that there appears another type of thermodynamic instability, 
which is subsequently analyzed by means of the free energy.  
%
%
%
%
\subsection{Thermodynamic temperature from the Clausius relation}
\label{subsec: temperature}
%
%
%
%
%
As has been mentioned in section \ref{sec: intro}, the concept of 
temperature is non-trivial in non-extensive thermostatistics. This is 
because the standard framework of thermodynamics crucially depends on the 
assumption of extensivity of entropy. According to the recent claim, the 
definition of physical 
temperature $\Tphys$ should be altered depending on the choice of 
energy constraint and is related to the inverse of the 
Lagrange multiplier, $1/\beta$, with {\it some correction factors} 
\cite{MNPP2000,AMPP2001}.   
Note, however, that this discussion heavily relies on the extensivity 
of the energy as well as the thermodynamic zeroth law. 
In our present case, the maximum entropy principle was applied subject 
to the constraints $E$ and $M$, adopting the standard definition of mean 
values (see eqs.(\ref{eq: energy})(\ref{eq: mass})). As a consequence, 
the resultant energy $E$ becomes non-extensive and we cannot apply 
the above definition.

To address the physical temperature in the present case, 
we therefore consider the relation between 
the heat transfer and entropy change and seek the most plausible candidate 
for thermodynamic temperature. That is, we analyze the 
variation of equilibrium configuration under fixing the total mass. 
Specifically, we deal with the quasi-static variation along an 
equilibrium sequence.

Let us first write down the heat change. The thermodynamic 
first law states that 
\begin{equation}
  \label{eq: d'Q}
  d'Q\,\, = \,\, dE \,+\, P_e\,dV,
\end{equation}
where the operation $d'$ stands for incomplete differentiation. 
The subscript $_e$ denotes a quantity evaluated at the edge.  
In the spherically symmetric configuration, the second term in 
right-hand side of (\ref{eq: d'Q}) becomes $4\pi r_e^{2} P_e\,dr_e$. 
As for the first term,  
the energy of the stellar polytropic system within the radius $r_e$, 
is computed in Appendix A.1. Introducing the dimensionless parameter 
$\lambda$, it is expressed in terms of the homology invariants 
as follows: 
\begin{eqnarray}
  \lambda &\equiv& -\frac{r_eE}{GM^2}=
\,-\frac{1}{n-5}\,
\left[\,\frac{3}{2}\left\{1-(n+1)\frac{1}{v_e}\right\}
  +(n-2)\frac{u_e}{v_e}\,\right], 
\label{eq: lambda}
\end{eqnarray}
where the quantity with subscript $_{e}$ represents the one evaluated 
at the boundary $r=r_e$. 
Using (\ref{eq: lambda}),  the heat change $d'Q$ is rewritten as follows: 
\begin{eqnarray}
  &&  d'Q\,=\,d\left(-\lambda\,\frac{GM^2}{r_e}\right)\,
+\,4\pi r_e^2\,P_e\,dr_e,
\nonumber \\
&&~~~~~~~~~~~~~~~
=\,\frac{GM^2}{r_e}\,\,\left\{\left(\lambda+\frac{u_e}{v_e}\right)\,\,
\frac{dr_e}{r_e}\,-\,\xi_e\,\frac{d\lambda}{d\xi_e}\,\,
\frac{d\xi_e}{\xi_e}\right\}, 
\label{eq: d'Q_2}
\end{eqnarray}
where the relation $4\pi r_e^4 P_e/(GM^2)=u_e/v_e$ is used in the last 
line (see definitions (\ref{eq: def_u})(\ref{eq: def_v})). 
In the above expression, derivative of $\lambda$ with respect to $\xi_e$ 
can be computed with a help of relation (\ref{eq: d(u,v)/dxi}) (see eq.(33) 
of ref.\cite{TS2002}): 
\begin{equation}
  \xi_e \frac{d\lambda}{d\xi_e}\, =\,\frac{n-2}{n-5}
\,\,\frac{g(u_e,\,v_e)}{2v_e} ,
\label{eq: d_lambda/d_xi}
\end{equation}
where 
\begin{equation}
 g(u,v)=4u^2+2uv-\left\{8+3\left(\frac{n+1}{n-2}\right)\right\}u-
\frac{3}{n-2}\,v+3\left(\frac{n+2}{n-2}\right).
\label{eq: g(u,v)}
\end{equation}

Next focus on the change of the entropy. 
From (\ref{Appen_A: entropy_uv}) 
in Appendix A.2, the entropy of the extremum state is given by 
\begin{equation}
  \label{eq: entropy_homology_re}
  S_q = \left(n-\frac{3}{2}\right)
\left[\,\frac{1}{n-5}\,\,\frac{\beta GM^2}{r_e}\,
\left\{2\frac{u_e}{v_e}-(n+1)\frac{1}{v_e}+1
        \right\} +N\right].
\end{equation}
Hence, the variation of entropy $dS_q$ under fixing the total mass 
can be decomposed into the variation of homology 
invariants $(u_e,v_e)$, radius $r_e$ and 
Lagrange multiplier $\beta$ as follows:
\begin{eqnarray}
  dS_q &=& \frac{n-3/2}{n-5}\,\,\frac{\beta GM^2}{r_e}\,
\left[\left(\frac{d\beta}{\beta}-\frac{dr_e}{r_e}\right)\,
    \left\{2\frac{u_e}{v_e}-(n+1)\frac{1}{v_e}+1\right\}
\right.\nonumber\\
&&~~~~~~~~~~~~~~~~~~~~~~~~~ \left.
+\left\{2\frac{u_e}{v_e}
    \left(\frac{du_e}{u_e}-\frac{dv_e}{v_e}\right)
    -\frac{n+1}{v_e}\frac{dv_e}{v_e}\right\}   
\right].
  \label{eq: dS_q_1}
\end{eqnarray}
Among these variations, variation of homology invariants is simply 
rewritten with $d\xi_e$, through the relation (\ref{eq: d(u,v)/dxi}). 
On the other hand, from the mass conservation, the variation of Lagrange 
multiplier, $d\beta$ is related to both the variations of homology 
invariants and $dr_e$ as follows. Using the condition of hydrostatic 
equilibrium at the edge $r_e$, 
one can obtain the following relation (see derivation in Appendix A.3): 
\begin{equation}
\label{eq: eta}
\eta\equiv
\left\{\frac{(GM)^{n}(\unitm\beta)^{n-3/2}}{r_e^{n-3}h^{3}}\right\}^{1/(n-1)}
\,\,=\,\,\alpha_n\,\left(u_e\,v_e^n\right)^{1/(n-1)}, 
\end{equation}
where the constant $\alpha_n$ is given by 
\begin{equation}
\label{eq: kappa_n}
\alpha_n \,\,=\,\, \left\{\frac{(n-1/2)^{n-3/2}}{16\sqrt{2}\pi^2\,
(n+1)^n\,B(3/2,n-1/2)}\right\}^{1/(n-1)}, 
\end{equation}
which asymptotically approaches unity, in the limit $n\to+\infty$. 
Keeping the total mass $M$ constant, variation of 
(\ref{eq: eta}) yields
\begin{equation}
\frac{n-3/2}{n-1}\,\,\frac{d\beta}{\beta}-\frac{n-3}{n-1}\,\,\frac{dr_e}{r_e}
=\frac{1}{n-1}\left(\frac{du_e}{u_e}+ n\,\,\frac{dv_e}{v_e}\right). 
\end{equation}
We then rewrite it with 
\begin{equation}
  \label{eq: beta_re}
  \frac{d\beta}{\beta}-\frac{dr_e}{r_e}=\frac{1}{n-3/2}\,\,
\left(\,-\frac{3}{2}\frac{dr_e}{r_e}+\,\frac{du_e}{u_e}\,+\,n\,\frac{dv_e}{v_e}
 \,\right).
\end{equation}
Substituting the relation (\ref{eq: beta_re}) into 
equation (\ref{eq: dS_q_1}), the dependence of $d\beta/\beta$ can be 
eliminated. 
Thus, using the relation (\ref{eq: d(u,v)/dxi}), the final form of the 
entropy change is expressed in terms of the variations $d\xi_e$ 
and $dr_e$. After some manipulation, we obtain
\begin{eqnarray}
&&  dS_q = \frac{\beta GM^2}{r_e}\,\left[\,
  -\frac{3/2}{n-5}\left(2\,\frac{u_e}{v_e}-\frac{n+1}{v_e}+1\right)
  \frac{dr_e}{r_e} -\frac{n-2}{n-5}\,\frac{1}{2v_e}
\right.
\nonumber\\
&&\times  \left.\left\{
  4u_e^2+2u_e v_e -\left(8+3\frac{n+1}{n-2}\right)u_e-
  \frac{3}{n-2}\,v_e +3\left(\frac{n+1}{n-2}\right)\right\}
\frac{d\xi_e}{\xi_e}
\right].
  \label{eq: dS_q_2}
\end{eqnarray}

Now, from the knowledge of the expressions $\lambda$ and 
$\xi_e(d\lambda/d\xi_e)$, one can easily show that the above equation 
is just identical to 
  \begin{equation}
    \label{eq: dS_q_3}
    dS_q = \frac{\beta GM^2}{r_e}\,\,\left\{\left(\lambda+\frac{u_e}{v_e}\right)
    \frac{dr_e}{r_e}\,-\xi_e\,\frac{d\lambda}{d\xi_e}\,\,
    \frac{d\xi_e}{\xi_e}\,\right\}.  
  \end{equation}
Therefore, comparison between (\ref{eq: dS_q_3}) and (\ref{eq: d'Q_2}) 
immediately leads to the following relation: 
\begin{equation}
\label{eq: Clausius_eq} 
        dS_q \,\,=\, \beta\,d'Q=\beta\,(dE + P_e dV),  
\end{equation}
which exactly coincides with the standard result 
of {\it Clausius relation} in a quasi-static process.

The relation (\ref{eq: Clausius_eq}) strongly suggests that the 
thermodynamic temperature $\Tphys$ is identified with the inverse of 
Lagrange multiplier, $\Tphys=1/\beta$. At first glance, 
the result seems somewhat trivial, since one can easily expect this 
relation from the standard thermodynamic relation, 
$\partial S_q/\partial E=\beta$, which generally holds even in the 
the non-extensive Tsallis formalism \cite{CT1991,PP1997}. 
As advocated by many author, however, the relation 
$\partial S_q/\partial E=\beta$ does not simply imply the thermodynamic 
temperature $\Tphys=1/\beta$ and it might even contradict with 
the thermodynamic temperature defined through the 
thermodynamic zeroth low \cite{AMPP2001}.

On the other hand, in our case of the self-gravitating system, 
the thermodynamic temperature $\Tphys=1/\beta$ is mathematically verified 
by the integrable condition of the thermodynamic entropy 
through the Clausius relation. 
Further, it is remarkably found that the relation $\Tphys=1/\beta$ 
holds even in the absence of gravity (the limit $G\to0$) and can be proven 
through an alternative route. In Appendix B, as a pedagogical example, we 
demonstrate that the relation $\Tphys=1/\beta$ is indeed obtained in the 
classical gas model using the Carnot cycle. 
%
%
%
%
\subsection{Negative specific heat and thermodynamic instability}
\label{subsec: specific heat}
%
%
%
%
%
%
Once obtained the thermodynamic temperature, $\Tphys=1/\beta$, 
we are in a position to investigate the thermodynamic instability from 
the straightforward calculation of the specific heat. 
Let us first discuss the qualitative behavior of the specific heat. 
By definition, the specific heat at constant volume is given by 
\begin{equation}
\label{eq: C_V_1}
        \Cv \equiv
        \left(\frac{dE}{d\Tphys}\right)_e 
=       -\beta^2\,\left(\frac{dE}{d\beta}\right)_e
= - \beta^2\,\,\frac{\displaystyle \left(\frac{dE}{d\xi}\right)_e}
        {\displaystyle \left(\frac{d\beta}{d\xi}\right)_e}.  
\end{equation}
Recall that the dimensionless parameters $\lambda$ and $\eta$ 
are respectively proportional to $-E$ and $\beta^{(n-1)/(n-3/2)}$ (see 
eqs.(\ref{eq: lambda})(\ref{eq: eta})). This implies that 
for a system of constant mass inside a fixed wall, the 
qualitative behavior of (\ref{eq: C_V_1}) can be deduced from the 
relation between $\eta$ and $\lambda$.

\begin{figure}[ht]
  \begin{center}
    \includegraphics*[width=10cm]{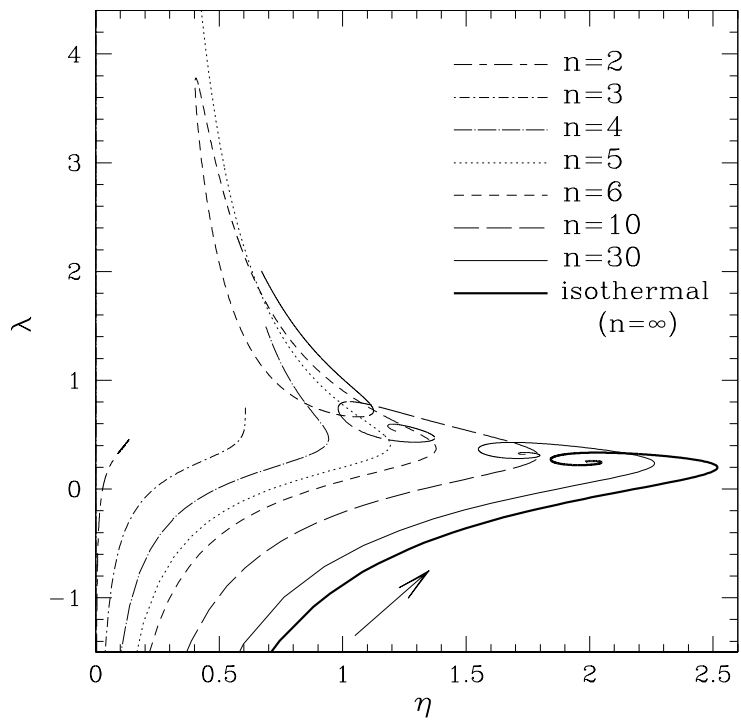}
  \end{center}
    \caption{Trajectory of Emden solutions in $(\eta,\lambda)$-plane.}
    \label{fig: eta_lambda}
\end{figure}

Figure \ref{fig: eta_lambda} depicts the trajectories of the Emden 
solutions in the $(\eta,\lambda)$-plane with various polytrope 
indices. Each point along the trajectory represents an Emden 
solution for different value of the radius $r_e$. 
From the boundary condition, all the trajectories start from 
$(\eta,\lambda)=(0,-\infty)$, corresponding to the origin $r_e=0$. 
As gradually increasing the radius, the trajectories first move to 
upper-right direction monotonically, as marked by the arrow. 
At this stage, the kinematic energy dominates the potential energy and 
the system lies in a kinematically thermal state ($\lambda<0$), 
indicating the positive specific heat. 
For larger radius, while the curves with index $n\leq3$ abruptly terminate,  
the trajectories with $n>3$ suddenly change their direction from upper-right 
to upper-left. Moreover,  in the case of $n>5$, the trajectory 
progressively changes its direction and it finally spirals around a 
fixed point.

From these observations,  one can roughly infer the existence of the 
two types of the thermodynamic instability as follows. At first inflection 
point for $n>3$, the specific heat diverges and the signature of $\Cv$ 
becomes indefinite. Beyond this point, the specific heat changes from 
positive to negative.  This means that the potential energy conversely 
dominates the kinetic energy,  indicating the system 
being {\it gravothermal}. In this case, equilibrium state ceases to exist 
for a system in contact with a heat bath, but does still exist for a system 
surrounded by an adiabatic wall. However, for the polytrope 
index $n>5$, the specific heat of the system turns to increase beyond this 
inflection point and it next reaches at the point $d\lambda/d\eta=0$, i.e, 
$\Cv=0$. This means that while the inner part of the system still keeps the 
specific heat negative, the fraction of the outer normal part grows up as 
increasing $r_e$ and it eventually balances with inner gravothermal part. 
Thus, beyond this critical point, no thermal balance is attainable and the 
system becomes gravothermally unstable. This is true even in the system 
surrounded by an adiabatic wall.

Now, let us write down the explicit expression for the specific heat 
$\Cv$. In equation (\ref{eq: C_V_1}), the variation of $\beta$ and $E$ 
with $\xi_e$ can be respectively rewritten with 
\begin{equation}
\left(\frac{dE}{d\xi}\right)_e = \,-\,\,\frac{GM^2}{r_e}\,\,
        \frac{d\lambda}{d\xi_e},
\label{eq: dEdxi}
\end{equation}
and
\begin{equation}
\left(\frac{d\beta}{d\xi}\right)_e = \,\frac{n-1}{n-3/2}\,
        \frac{\beta}{\eta}\,\,\frac{d\eta}{d\xi_e}. 
\label{eq: detadxi}
\end{equation}
Here, the variable 
$d\lambda/d\xi_{e}$ has been already given in (\ref{eq: d_lambda/d_xi}). 
As for the derivative of $\eta$ with respect to $\xi_e$, we obtain 
\begin{equation}
  \label{eq: d_eta/d_xi}
\xi_e\,\,\frac{d\eta}{\xi_e}\,\,=\,\,
\left(u_{e}-\frac{n-3}{n-1}\right)\,\eta.   
\end{equation}
Then the quantity $\Cv$ becomes 
\begin{eqnarray}
      \Cv \,\,=\,\,\frac{(n-3/2)(n-2)}{(n-1)(n-5)}\,\,
        \frac{\beta GM^2}{r_e}\,\frac{g(u_e,v_e)}{\displaystyle 
          2v_e\,\left(u_e-\frac{n-3}{n-1}\right)}, 
\nonumber 
\end{eqnarray}
with the function $g(u_e,v_e)$ given by (\ref{eq: g(u,v)}). 
Notice that the above expression is still redundant, since there remains 
the explicit dependence of the variable $\beta$. Eliminating the variable 
$\beta$ by using the relation (\ref{eq: eta}), one finally obtains
\begin{equation}
  \label{eq: C_v_2}
\frac{\Cv}{N} = \tilde{\alpha}_n\,\left(\frac{h^2}{GMr_e}\right)^{(3/2)/(n-3/2)}
\frac{g(u_e,v_e)}{\displaystyle 2 \left(u_e-\frac{n-3}{n-1}\right)}\,\,
\left(u_e\,\, v_e^{3/2}\right)^{1/(n-3/2)},  
\end{equation}
where we introduced the new dimensionless constant $\tilde{\alpha}_n$: 
\begin{equation}
  \label{eq: tilde_alpha_n}
  \tilde{\alpha}_n \equiv \frac{(n-3/2)(n-2)}{(n-1)(n-5)}\,\,
\alpha_n^{1/(n-3/2)}. 
\end{equation}
Note that in the limit $n\to+\infty$, equation (\ref{eq: C_v_2}) 
consistently recovers the well-known result of isothermal sphere 
(e.g, eq.(39) of ref.\cite{Chavanis2002}):
\begin{equation}
  \label{eq: C_v_iso}
  \frac{\Cv}{N}\,\,\,\stackrel{n\to+\infty}{\longrightarrow}\,\,\,
  \frac{4u_e^2 + 2u_e v_e -11 u_e +3}{2(u_e-1)}.
\end{equation}
Comparing (\ref{eq: C_v_2}) with the isothermal limit, the resultant 
expression contains a residual dimensional parameter 
$h$, as well as the quantities $M$ and $r_e$. While the residual 
dependence can be regarded as a natural consequence of the 
non-extensive generalization of the entropy, it would be helpful 
to understand the origin of this scaling in more simplified manner.  
This will be discussed in section \ref{sec: discussion}.

Apart from the residual factor, the expression of specific heat 
(\ref{eq: C_v_2}) clearly reveals the two types of thermodynamic instability 
seen in Figure \ref{fig: eta_lambda}. The inflection point with the infinite 
specific heat, $\Cv\to\pm\infty$ leads to the condition 
\begin{equation}
  u_e -  \frac{n-3}{n-1}=0,       
  \label{eq: criterion_1}
\end{equation}
which immediately yields the conclusion that this is only possible for the 
polytrope index $n>3$, consistent with Figure \ref{fig: eta_lambda}. 
On the other hand, critical point with the vanishing specific heat, 
$\Cv=0$ corresponds to the following condition:  
\begin{equation}
  g(u_e,\,v_e)=0.  
  \label{eq: criterion_2}
\end{equation}
This is exactly the same condition as obtained from the second variation 
of entropy (see eq.(33) or (53) in ref.\cite{TS2002}). According to the 
previous analysis, the condition (\ref{eq: criterion_2}) represents the 
marginal stability at which the extremum state of the entropy $S_q$ is 
neither maximum nor minimum. This situation turns out to appear when 
the polytrope index $n>5$.

Therefore, we reach a fully satisfactory conclusion that the 
thermodynamic instability found from the second variation of entropy is 
intimately related to the presence of negative specific heat and the 
stability/instability criterion can be 
exactly recovered from the critical point of the thermal balance, $\Cv=0$, 
which is also consistent with the analysis in the Boltzmann-Gibbs limit, 
$n\to\infty$ \cite{LW1968}. The successful result can be regarded as an 
outcome of the correct definition of $\Tphys$. 
As for the transition point with $\Cv\to\pm\infty$, it clearly  
indicates the thermodynamic instability of a system in contact with a 
thermal bath. In next section, by means of the free energy, we confirm that 
the condition (\ref{eq: criterion_1}) indeed represents the marginal 
stability of the system surrounded by a thermal wall and 
beyond this point the system will be unstable.

\begin{figure}
  \begin{center}
    \includegraphics*[width=13.cm]{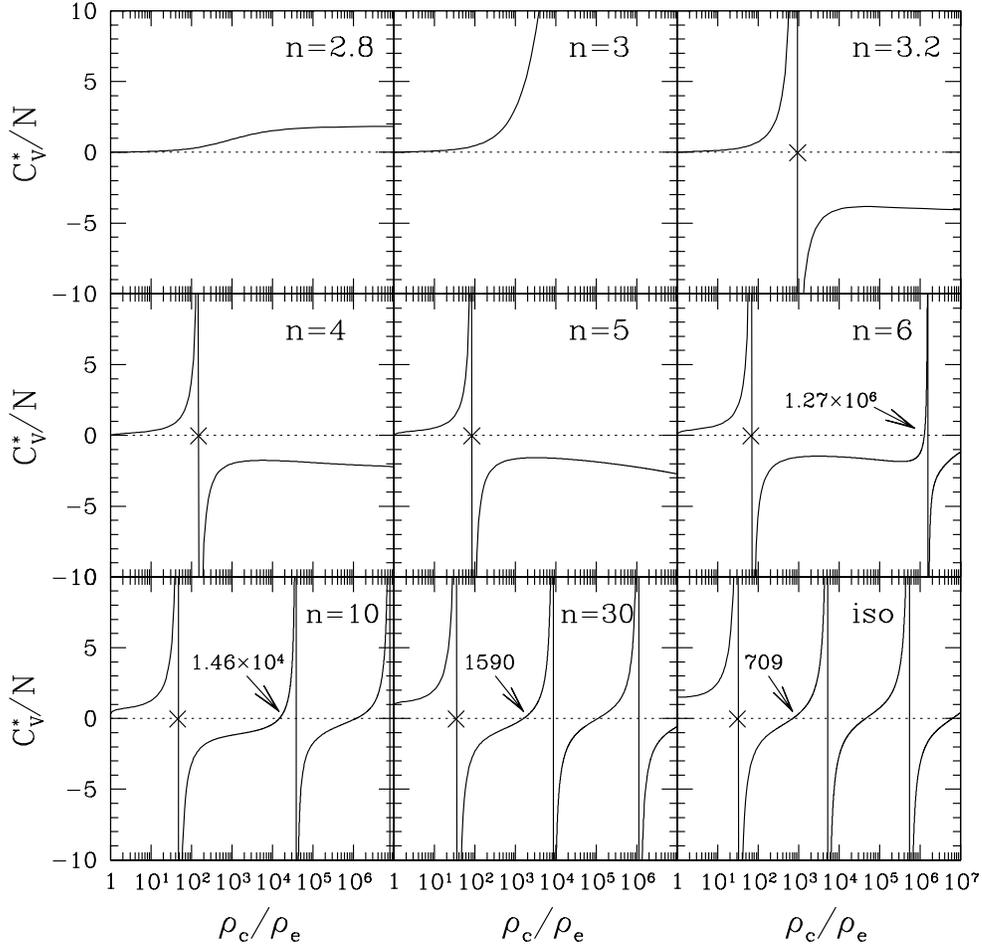}
  \end{center}
    \caption{Normalized specific heat per particle
      $C_{\rm \scriptscriptstyle V}^*/N$ 
      as a function of density contrast 
      $\rho_c/\rho_e$ near the critical polytrope indices 
        $n=3$({\it upper}) and $n=5$({\it middle}), and 
        large $n$ cases({\it lower}). Here, 
        the normalized specific heat $\Cv^*$ is defined by 
        $\Cv/(h^2/GMr_e)^{(3/2)/(n-3/2)}$. 
        \label{fig: C_V}}
\end{figure}

In Figure \ref{fig: C_V}, by varying the radius $r_e$, 
the normalized specific heat per particle 
$\Cv^{*}/N$ is plotted as a function of density contrast, 
$\rho_c/\rho_e$ 
around the critical polytrope indices $n=3$({\it upper-panels}) and 
$n=5$({\it middle-panels}). Here, the normalized specific heat $\Cv^{*}$ is defined 
by the specific heat $\Cv$ divided by the redundant factor 
$(h^2/GMr_e)^{(3/2)/(n-3/2)}$. Obviously, the transition point 
$\Cv\to\pm\infty$ appears when $n>3$({\it crosses}), 
while the existence of critical point $\Cv=0$ is allowed for 
higher density contrast of $n>5$ cases({\it arrows}). 
The critical values $D_{\rm crit}\equiv(\rho_c/\rho_e)_{\rm crit}$ 
indicated by arrows exactly coincide
with those obtained from the previous analysis (see Table 1 of 
ref.\cite{TS2002}). Lower-panels of Figure \ref{fig: C_V} show the 
specific heat with large polytrope indices $n=10$ and $30$, together with 
the Boltzmann-Gibbs limit ($n\to+\infty$, labeled by {\it iso}). As increasing 
the polytrope 
index $n$, the critical/transition points tend to shift to the lower density 
contrast, while the successive divergent and zero-crossing points appear 
at the higher density contrast, corresponding to the behavior seen in 
Figure \ref{fig: eta_lambda}. 
%
%
%
%
%
%
%
%
%
%
%
%
\section{Thermodynamic instability from the second variation of free energy}
\label{sec: free-energy}
%
%
%
%
%
%
%
Previous section reveals that there exists another type of 
thermodynamic instability in which 
the marginal stability is deduced from the 
condition (\ref{eq: criterion_1}).  
In this section, to check the consistency of the non-extensive 
thermostatistics, we reconsider this issue by means of the Helmholtz 
free energy: 
\begin{equation}
\label{eq: free_energy} 
        F_q \,=\, E \,-\, \Tphys\,S_q. 
\end{equation}
Adopting the relation $\Tphys=1/\beta$,  
we re-derive the marginal stability condition (\ref{eq: criterion_1}) 
from the second variation of $F_q$.

Consider a system surrounded by the thermally conducting wall 
in contact with a heat bath. Usually, the stable equilibrium state 
should keep the free energy $F_q$ minimum.  
Thus the presence of thermodynamic instability implies the absence of 
minimum free energy, which can be deduced from the signature of the second 
variation $\delta^2F_q$ around the extremum state of free energy. 
Since the non-extensive formalism still verifies the Legendre 
transform structure leading to the standard result of thermodynamic 
relation\cite{CT1991,PP1997}, the extremum state of the free energy 
exactly coincides with that of the entropy. One thus skips to find the 
extremum state of $F_q$ and proceeds to evaluate the second order variation.

In contrast to the adiabatic treatment, we here deal with the density 
perturbation $\rho\to\rho+\delta\rho$, surrounded by a thermal wall. 
To be specific, we evaluate the second variation 
under keeping the radius $r_e$, the total mass $M$ and the 
temperature $\Tphys$ 
constant. Then the variation of energy up to the second order leads to 
\begin{eqnarray}
\delta E&=& \,\delta\left[\int\left\{\frac{3}{2}\,P(x)+\frac{1}{2}
\rho(x)\Phi(x)\right\} d^3\xx\,\right],
\nonumber \\
\nonumber \\
&=& \int \left\{\frac{3}{2}\,\delta P + 
\frac{1}{2}(\delta\rho\,\Phi+\rho\,\delta\Phi) +\frac{1}{2}\,\delta\rho\,\delta\Phi\,
\right\} d^3\xx.
\label{eq: delta_E}
\end{eqnarray}
Similarly, using the expression (\ref{Appen_A: tsallis_entropy}) 
in Appendix A.2, the variation of Tsallis entropy becomes 
\begin{eqnarray}
\delta S_q &=& \,\delta\,\left[
\left(n-\frac{3}{2}\right)\,\left\{N-\beta\,\int\,\,P(x)\,d^3\xx\,\right\}\,
\right],
\nonumber\\
\nonumber\\
&=& -\,\left(n-\frac{3}{2}\right) \, \beta\,\int\, \delta\,P(x) d^3\xx.
\label{eq: delta_S_q}
\end{eqnarray}
The above expressions include the variation of pressure $\delta P$,    
which can be expanded with a help of the polytropic equation of state 
(\ref{eq: Eq. of state}):  
\begin{equation}
  \delta P= \left(1+\frac{1}{n}\right)\,\frac{P}{\rho}\,\delta\rho + 
  \frac{1}{2}\,\left(1+\frac{1}{n}\right)\frac{1}{n}\,\frac{P}{\rho^2}\,
  \left(\delta\rho\right)^2. 
\label{eq: delta_P}
\end{equation}
Combining the above result with equations (\ref{eq: delta_E}) and 
(\ref{eq: delta_S_q}) and collecting the second order terms only, the second 
variation of free energy becomes
\begin{eqnarray}
\delta^2 F_q = \delta^2 E - \Tphys \,\delta^2 S_q
= \frac{1}{2}\,\int \left\{\,\frac{n+1}{n}\,\frac{P}{\rho^2}\,
  \left(\delta\rho\right)^2\,+\,\delta\rho\,\delta\Phi\,\right\}\,d^3\xx, 
\label{eq: del^2_F_q}
\end{eqnarray}
where the relation $\Tphys=1/\beta$ is used in the last line. 
Now, restricting our attention to the spherical symmetric perturbation, 
we introduce the following perturbed quantity (see refs.
\cite{TS2002}\cite{Padmanabhan1989}): 
\begin{equation}
\label{eq: Q(r)}        
\delta\rho(r)=\frac{1}{4\pi r^2}\,\frac{dQ(r)}{dr}.
\end{equation}
Then the mass conservation $\delta M=0$ implies the boundary condition 
$Q(0)=Q(r_e)=0$. Substituting (\ref{eq: Q(r)}) into (\ref{eq: del^2_F_q}) and 
repeating the integration by part, one finally reaches the following 
quadratic form: 
\begin{eqnarray}
\label{eq: d^2_F}       
        \delta^2F_q \,=\, -\frac{1}{2}\,\,\int_0^{r_e}dr\,\,
        Q(r)\left[\,\frac{n+1}{n}\,\frac{d}{dr}\left\{\frac{1}{4\pi\,r^2\,\rho}
        \,\left(\frac{P}{\rho}\right) \frac{d}{dr}\right\}+\frac{G}{r^2}
        \right]Q(r).
\end{eqnarray}

Thus, the problem just reduces to the eigenvalue problem and the 
stability of the system can be deduced from the signature of the eigenvalue.  
More specifically, the onset of instability corresponds to 
the marginally stability condition, $\delta^2F_q=0$, and it is 
sufficient to analyze the zero-eigenvalue equation: 
\begin{equation}
\label{eq: zero_eigenvalue}
        \hat{L}\,\,Q(r)\equiv
        \left[\frac{d}{dr}\left\{\frac{1}{4\pi\,r^2\,\rho}
        \,\left(\frac{P}{\rho}\right) \frac{d}{dr}\right\}+
        \frac{n}{n+1}\,\frac{G}{r^2}\right]\,Q(r)\,=\,0, 
\end{equation}
with the boundary condition, $Q(0)=Q(r_e)=0$. 
Equation (\ref{eq: zero_eigenvalue}) has quite similar form to the 
zero-eigenvalue equation found in the adiabatic treatment
(see eq.(46) of ref.\cite{TS2002}).  
Except for the non-local term, one can utilize the previous knowledge 
to solve the equation (\ref{eq: zero_eigenvalue}):
\begin{equation}
\hat{L}\,(4\pi r^3\,\rho) = \frac{n-3}{n+1}\,\frac{G\,m(r)}{r^2},~~~~~~~ 
\hat{L}\,m(r) = \frac{n-1}{n+1}\,\frac{G\,m(r)}{r^2}. 
\end{equation}
These two equation leads to the ansatz of the solution:   
\begin{equation}
\label{eq: analytic_solution}
Q(r) = c\left\{\,\,4\pi\,r^3\,\rho(r) -\frac{n-3}{n-1}\,\,m(r)\right\}. 
\end{equation}
Here, the variable $c$ is an arbitrary constant. The above equation 
(\ref{eq: analytic_solution}) automatically satisfies the boundary 
condition $Q(0)=0$, while the remaining 
condition $Q(r_e)=0$ puts the following constraint: 
\begin{equation}
\label{eq: stability_criterion}
Q(r_e)=c\left(4\pi\,r_e^3\,\rho_e-\frac{n-3}{n-1}\,M\right)=
c\left(\,u_e-\frac{n-3}{n-1}\,\right)M=0.     
\end{equation}
Again, we arrive at the satisfactory result that the solution of 
zero-eigenvalue equation exactly recovers the condition 
(\ref{eq: criterion_1}).

Now, remaining task is to show that the second variation 
$\delta^2 F_q$ becomes negative beyond the transition point of 
$\Cv\to\pm\infty$. One can rewrite the expression (\ref{eq: d^2_F}) with 
\begin{eqnarray}
  \delta^{2}F_q = \frac{1}{2}\,\left(H-1\right)\,\int_0^{r_e}\,\,
  \frac{GQ^2}{r^2}\,dr, 
\nonumber 
\end{eqnarray}
with the constant $H$ given by
\begin{equation}
  H\,\equiv\,\frac{\displaystyle \frac{n+1}{n}\,
\int_0^{r_e}\,\frac{1}{4\pi r^2 \rho}\left(\frac{P}{\rho}\right)
\left(\frac{dQ}{dr}\right)^2\,dr}
{\displaystyle \int_0^{r_e}\,\frac{GQ^2}{r^2}\,dr}. 
\label{eq: def_of_H}
\end{equation}
That is, the condition $H>1$ implies stable local minimum 
state of free energy, while the inequality $H<1$ represents unstable 
local maximum state. Integrating by part, equation (\ref{eq: def_of_H}) 
can be regarded as an eigenvalue equation with eigenvalue, $H$:  
\begin{equation}
 -\frac{d}{dr}\left\{\frac{1}{4\pi r^2\rho}\left(\frac{P}{\rho}\right)
\frac{dQ}{dr}\right\}  =H\,\,\frac{n}{n+1}\,\,\frac{GQ}{r^2}. 
\label{eq: eigensystem_H}
\end{equation}
Obviously, equation (\ref{eq: analytic_solution}) becomes the 
solution of above equation with the minimum eigenvalue, $H_{\rm min}=1$, 
if the condition (\ref{eq: stability_criterion}) is fulfilled.  
In this case, solution (\ref{eq: analytic_solution}) can be regarded as 
the ground state of the eigensystem (\ref{eq: eigensystem_H}), 
since the function (\ref{eq: analytic_solution}) does not possess any 
nodes between $[0,r_e]$. Therefore, for a suitably smaller radius 
$r_e$ or a smaller density contrast $\rho_e/\rho_c$ below the transition point, 
the eigenvalue $H$ should be larger than unity. Conversely, from continuity, 
the condition $H<1$  must be satisfied beyond the critical radius.

\begin{figure}
  \begin{center}
    \includegraphics*[width=13cm]{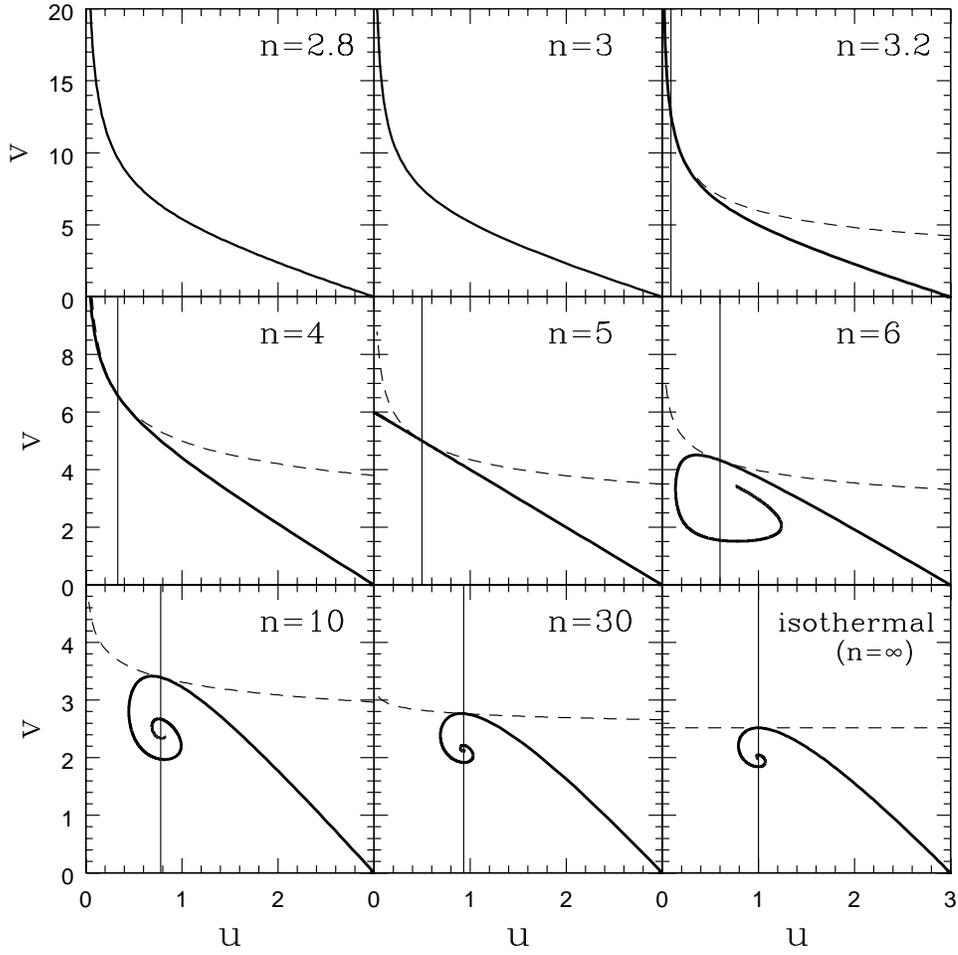}
  \end{center}
    \caption{Stability/instability criterion for a system 
        in contact with a thermal bath in $(u,v)$-plane. 
        The thick solid lines represent the trajectories of 
        Emden solutions, while the thin-solid and dashed lines 
        respectively denote the conditions (\ref{eq: criterion_1}) 
        and (\ref{eq: crit_curve}). 
      \label{fig: uv_crit}}
\end{figure}

Finally, using the $(u,v)$-variables, 
the geometrical meaning of onset of thermodynamic instability 
is briefly discussed in similar manner to the adiabatic case. 
In Figure \ref{fig: uv_crit}, the thick solid lines show the Emden 
trajectories with various polytrope indices in $(u,v)$-plane. 
The thin-solid lines in Figure 
\ref{fig: uv_crit} represents the straight lines, $u-(n-3)/(n-1)=0$. 
Since the equilibrium state only exists along the Emden trajectory, 
the condition (\ref{eq: stability_criterion}) is satisfied 
at the intersection of these two solid lines, which is only 
possible for $n>3$. 
On the other hand, as seen in previous section, the equilibrium system 
surrounded by a thermal wall is characterized by the 
three parameters, $r_e$, $M$ and $\beta$(or $\Tphys$), through the relation 
(\ref{eq: eta}). In other words, the system must lie on the curve: 
\begin{equation}
v=\left(\frac{\eta}{\alpha_n}\right)^{(n-1)/n}\,u^{-1/n}, 
\label{eq: crit_curve}
\end{equation}
with some constant value $\eta$. 
We have seen in Figure \ref{fig: eta_lambda} that the constant value 
$\eta$ is bounded from above, $\eta\leq\eta_{\rm crit}$. Thus, the 
critical curve (\ref{eq: crit_curve}) with $\eta=\eta_{\rm crit}$ must 
intersect with both the Emden trajectory and the straight line 
$u-(n-3)/(n-1)=0$ simultaneously. This is clearly shown in Figure 
\ref{fig: uv_crit}, 
where the critical curve is plotted as dashed lines. Since the critical 
curves tangentially intersect with Emden solutions, 
it always satisfies the condition $d\eta/d\xi=0$ at the contact point, 
leading to the condition (\ref{eq: criterion_1}) consistently.

Table 1 summarizes the dimensionless quantities $\eta_{\rm crit}$ and 
$D_{\rm crit}\equiv(\rho_{c}/\rho_{e})_{\rm crit}$ evaluated at the contact point. 
As increasing the polytrope index $n$, these values asymptotically approach 
the well-known results of Boltzmann-Gibbs limit, 
$\eta_{\rm crit}\to2.52$ and $D_{\rm crit}\to32.1$.
%
%
%
%
%
%
%
%
\section{Origin of non-extensive nature in stellar polytrope}
\label{sec: discussion}
%
%
%
%
%
%
%
%
%
As has been mentioned in section \ref{subsec: specific heat}, 
specific heat of the stellar polytropic system explicitly depends 
on the residual dimensional parameter $h$,  
in contrast to the isothermal limit (\ref{eq: C_v_iso}). 
In this section, to contact the physical meaning of the non-extensivity 
in stellar polytrope, we discuss the origin of this residual dependence. 
Indeed, the appearance of the residual factor can be recognized as the 
breakdown of 
both the intensivity of temperature and the extensivity of energy and entropy 
as follows. From equation (\ref{eq: Lane-emden_eq}), 
the asymptotic behavior of the Emden solution becomes  
\begin{eqnarray}
\theta \sim \xi^{-2/(n-1)},~~~~\rho \sim r^{-2n/(n-1)},
~~~~(\xi, r \rightarrow \infty)
\nonumber
\end{eqnarray}
so that the mass within a sphere of radius $r$ is given by 
\begin{equation}
M \sim \rho\, r^3 \propto r_e^{(n-3)/(n-1)}~~.
\label{eq: M-r}
\end{equation}
Then the energy of a virialized stellar system is roughly estimated as 
\begin{eqnarray}
E \sim \frac{GM^2}{r_e} \propto r_e^{(n-5)/(n-1)} \propto M^{(n-5)/(n-3)},
\nonumber
\end{eqnarray}
and the relation (\ref{eq: eta}) tells 
\begin{eqnarray}
\beta \propto r_e^{-(n-3)/(n-1)/(n-3/2)} \propto M^{-1/(n-3/2)}.
\nonumber
\end{eqnarray}
These relations clearly show the breakdown of the intensivity of 
temperature and the extensivity of energy, which lead to 
the scaling of the specific heat per mass: 
\begin{eqnarray}
\frac{C_{\rm V}}{N} = \frac{1}{M} \frac{dE}{d\Tphys} \sim \frac{\beta E}{M}
\propto  M^{-3(n-2)/(n-3)/(n-3/2)}~~. 
\label{eq: C_v-M}
\end{eqnarray} 
On the other hand, the dimensionless combination $h^2/(GMr_e)$ 
represents the ratio of a typical scale of the stellar system,  
$GMr \sim (GM/r) r^2 \sim v^2 r^2 $, to that of the reference cell,  
$h = v_0 \, l_0$. This behaves as
\begin{eqnarray}
\frac{h^2}{GMr_e}\, \propto\, \frac{1}{Mr_e}\, \propto\,  M^{2(n-2)/(n-3)}. 
\label{eq: h2/GMr-M}
\end{eqnarray}  
Thus, these two equations (\ref{eq: C_v-M}) and (\ref{eq: h2/GMr-M}) lead to the 
scaling relation of (\ref{eq: C_v_2}): 
\begin{equation}
\frac{C_{\rm V}}{N} \, \sim \,\, 
\left(\frac{h^2}{GMr_e}\right)^{(3/2)/(n-3/2)}.
\label{eq: scaling again}
\end{equation}
Notice that the Clausius relation (\ref{eq: Clausius_eq}) 
suggests that the entropy per unit mass has the same scaling relation: 
\begin{eqnarray}
\frac{S_q}{M} \sim \frac{\beta E}{M} \sim \frac{C_{\rm V}}{N},
\nonumber
\end{eqnarray}
Therefore, resultant dependence (\ref{eq: scaling again}) for the stellar 
polytrope can be a natural outcome of the non-extensivity of the entropy.

In fact, framework of the thermostatistics generally requires an 
introduction of the scale of the unit cell in order to count the available 
number of states in phase spaces. This is even true in the case of the 
isothermal stellar system($n\to+\infty$ or $q\to1$), but, 
the thermodynamic quantities show somewhat peculiar dependence of the scale 
$h$. A typical example is the entropy: 
\begin{eqnarray}
   S_{\rm BG} = \frac{M}{m_0} 
        \left\{ 
        \left(2 u_e+v_e-\frac{9}{2} \right) 
        - \ln\left(\frac{u_e v_e^{3/2}}{4\pi}\right)    
        - \frac{3}{2} \ln\left(\frac{h^2}{2\pi GMr_e}\right)
        \right\},
\nonumber
\end{eqnarray}
where $u_e$ and $v_e$ are the homology invariants for the isothermal system. 
The above equation shows that in the Boltzmann-Gibbs limit, 
$h$-dependence of the entropy can be recognized as a matter of choice of 
an additive constant, so that its derivatives, e.g., specific heat, is 
free from the residual dependence.

It should be emphasized that the stellar equilibrium 
system recovers the extensivity in the limit $n\to\infty$ and it behaves as  
\begin{eqnarray}
E \sim M \sim r,\,\,\, C_{\rm V} \sim M.
\label{eq: iso-case}
\end{eqnarray}
Also, the temperature becomes intensive in this limit. 
Thus, we readily understand that the 
scaling behavior shown in (\ref{eq: C_v_2}) or 
(\ref{eq: scaling again}) has nothing to do with the long-range nature 
of the gravity. Even in the free polytropic gas model in Appendix B, 
the residual dependence emerges as 
\begin{eqnarray}
\frac{C_{\rm V}}{N} \sim \,\, 
\left\{\frac{h^2}{(P/\rho) V^{2/3}}\right\}^{(3/2)/(n-3/2)}. 
\nonumber
\end{eqnarray}
It follows that the explicit dependence of the specific heat 
on the reference cell scale $h$ just originates from the  
the non-extensive nature of Tsallis entropy.  
%
%
%
%
%
%
%
%
\section{Summary \& Discussions}
\label{sec: summary}
%
%
%
%
%
%
%
%
%
In this paper, thermodynamic properties of 
the stellar self-gravitating system arising from Tsallis' non-extensive 
entropy have been studied in detail. In particular, physical interpretation 
of the thermodynamic instability previously found from the second 
variation of entropy is discussed in detail within a 
framework of the non-extensive thermostatistics. After briefly reviewing the 
equilibrium state of Tsallis entropy, we first address the issues on  
thermodynamic temperature in the case of equilibrium stellar polytrope. 
Analyzing the heat transfer and the entropy change in a quasi-static 
process, standard form of the Clausius relation is derived,  
irrespective of the non-extensivity of entropy. According to this result, 
we explicitly calculate the specific heat and confirm the presence of 
negative specific heat. The onset of instability found in previous work 
just corresponds to the 
zero-crossing point, $\Cv=0$, supporting the fact that the heuristic 
explanation of gravothermal catastrophe 
holds even in the non-extensive thermostatistics.

Further, the analysis of 
specific heat shows divergent behavior at $n>3$, 
suggesting another type of 
thermodynamic instability, which occurs when the system is surrounded 
by a thermal wall. We then turn to the stability analysis 
by means of the Helmholtz free energy. Similar to the previous early work, 
the stability/instability criterion just reduces to the solution of the 
zero-eigenvalue problem and solving the eigenvalue equation, 
we recover the marginal stability condition derived from the 
divergence of specific heat (\ref{eq: criterion_1}). 

In addition to the thermostatistic treatment, we have also discussed 
the origin of non-extensivity in stellar polytrope. The residual 
dependence of the reference scale $h$ appeared in the specific heat 
(\ref{eq: C_v_2}) naturally arises from the non-extensivity of the entropy 
and the resultant scaling dependence can be simply deduced from 
the asymptotic behavior of the Emden solutions.

The stability analysis using the free energy in section 
\ref{sec: free-energy} is consistent with recent claim by 
Chavanis \cite{Chavanis2001}, who has 
investigated the dynamical instability of polytropic gas sphere. 
According to his early paper \cite{Chavanis2002}, 
the thermodynamic stability of stellar system is intimately 
related to the dynamical stability of gaseous system, which has been 
clearly shown in the case of the isothermal distribution. 
Thus, the correspondence between Chavanis' recent result 
\cite{Chavanis2001} and a part of our present analysis 
can be regarded as a generalization of his early work 
to the polytropic system. Note, however, that starting from the Tsallis 
entropy, we extensively discuss the thermodynamic temperature 
and the specific heat of stellar polytrope. 
Therefore, at least, from the thermodynamic point of view, 
our present analysis provides a valuable insight to the stellar 
equilibrium systems.

A particular interest in the thermodynamic relation is 
the Clausius relation (\ref{eq: Clausius_eq}) that has been 
still preserved in the non-extensive stellar system. This is 
indeed consistent with 
the standard thermodynamic relation $\partial S/\partial E=\beta$,  
if one keeps the volume constant. Note also that the relation 
$\partial S/\partial E=\beta$ is readily obtained from 
the standard Legendre transform structure. While we only dealt with 
a specific case with the non-extensive entropy (\ref{eq: S_q}),  
it is well known 
that the standard Legendre transform structure does generally hold 
independently of the functional form of the entropy \cite{PP1997}.  
Hence, our result in turn suggests that the Clausius relation is also 
valid for any stellar system maximizing the entropic functional more  
general than Tsallis'.

At present, the results shown in this paper seems fully consistent with 
  the general framework of the thermostatics. Apart from the thermodynamic 
  instability, the stellar polytropic system can be a plausible thermodynamic 
  equilibrium state, as well as the isothermal stellar distribution. 
  In the isothermal case, existence of the thermodynamic limit has 
  been discussed by de Vega and S\'anchez \cite{deVegaSanchez}: 
\begin{eqnarray}
M,~V \rightarrow \infty, ~~~\frac{M}{V^{1/3}} = {\rm fixed},
\nonumber
\end{eqnarray}
 where $V \sim r^3$ is a volume of the system. Recalling the discussion 
 in section \ref{sec: discussion}, the above condition merely reflects 
 the extensivity of the isothermal system (\ref{eq: iso-case}). Thus, 
 similar argument can hold for the non-extensive system. 
 According to the scaling relation (\ref{eq: M-r}), 
 the existence of the thermodynamic limit in stellar polytrope yields 
 the condition: 
\begin{eqnarray}
  M,~V \rightarrow \infty, ~~~\frac{M}{V^{(n-3)/(3n-3)}} = {\rm fixed}.
  \nonumber
\end{eqnarray} 

Note, however, that this discussion relies on the 
non-uniqueness of the Boltzmann-Gibbs theory, which can be proven only 
mathematically\cite{AR2000}. Indeed, framework of the thermostatistics 
cannot answer the question whether the stellar polytropic distribution 
is really achieved as a thermodynamic equilibrium. To address this issue, 
we must study the detailed process of the long-term stellar dynamical 
evolution.  In the light of this, the analysis using Fokker-Planck model 
or direct N-body simulation can provide an invaluable insight to the 
non-extensive nature of stellar gravitating systems. This issue is now 
in progress and will be presented elsewhere.

Another remaining issue is the re-examination of the present analysis 
from a view of the 'standard' Tsallis formalism 
using the normalized $q$-expectation values. 
Apart from some technical issues on the treatment of the 
maximum entropy principle, one might naively expect that the consistency 
between the statistical and the thermodynamic analysis should be preserved 
even in the new formalism. However, a rather subtle point would be 
the identification of the thermodynamic temperature. As several author 
stated, the standard Clausius relation should be modified in the new 
Tsallis formalism and the resultant form of the expression apparently 
seems to contradict with the 
thermodynamic temperature defined through the thermodynamic zeroth law 
\cite{AMPP2001}. This point will be in particular important in discussing 
the thermodynamic instability and should be clarified along the line of 
our present treatment.

\bigskip
This work is supported in part by the Grant-in-Aid for Scientific Research 
of Japan Society for the Promotion of Science. 
\clearpage
%
%
%
%
%
\section*{Appendix A:  Thermodynamic variables in a stellar polytropic system}
\label{appen_A}
%
%
%
In this appendix, using the equilibrium state of stellar polytrope 
described in section \ref{sec: polytrope}, we explicitly evaluate the 
thermodynamic variables,  which have been used in section 
\ref{sec: non-extensive} and \ref{sec: free-energy}.
%
%
%
%
%
\subsection*{A.1~ Energy}
%
%
%
%
%
%
Recall that the equilibrium system confined in a spherical container 
satisfies the following virial theorem (e.g, p.502 of Ref.\cite{BT1987}): 
\begin{eqnarray}
  2K+U = 4\pi r_e^3\,P_e. 
\nonumber
\end{eqnarray}
The energy (\ref{eq: energy}) is then expressed as 
\begin{eqnarray}
  E = K+U = 4\pi r_e^3\,P_e - \,K 
    = 4\pi r_e^3\,P_e - \,\frac{3}{2}\int_{0}^{r_e}\,P(r)4\pi\,r^2 dr.
\end{eqnarray}
To evaluate the above integral in the 
spherically symmetric case, we use the following 
integral formula:  
\begin{equation}
\int_0^{r_e} P(r)\,4\pi r^2 dr =  -\frac{1}{n-5}\,
\left\{8\pi\,r_e^3\,P_e-(n+1)\,\frac{MP_e}{\rho_e}+
\frac{GM^2}{r_e}\right\}, 
\label{Appen_A: integral}
\end{equation}
which can be derived from the conditions of hydrostatic equilibrium,  
(\ref{eq: hydro_1}) and (\ref{eq: hydro_2}) 
(see Appendix A of ref.\cite{TS2002}). 
Thus, the energy of extremum state becomes 
\begin{eqnarray}
E  = \frac{1}{n-5}\,\left[\frac{3}{2} \left\{ \frac{GM^2}{r_e}-
        (n+1)\frac{MP_e}{\rho_e} \right\}+(n-2)\,4\pi r_e^3P_e\right].
\label{Appen_A: energy}
\end{eqnarray}
In terms of the homology invariants, we obtain 
\begin{equation}
  E =\frac{1}{n-5}\,\frac{GM^2}{r_e}\,
\left[\,\frac{3}{2}\left\{1-(n+1)\frac{1}{v_e}\right\}
  +(n-2)\frac{u_e}{v_e}\,\right].
\label{Appen_A: energy_uv}
\end{equation}
%
%
%
%
\subsection*{A.2~ Entropy}
%
%
%
%
First note the definition of Tsallis entropy (\ref{eq: S_q}): 
\begin{eqnarray}
  \nonumber
  S_q = -\left(n-\frac{3}{2}\right)\left\{\int\,
N\,\left(\frac{f}{N}\right)^{(n-1/2)/(n-3/2)}d^6\bftau 
- N \right\}.
\end{eqnarray}
Substituting the distribution function (\ref{eq: df_1}) into the above 
equation, after some manipulation, we obtain
\begin{eqnarray}
    S_q  = -\left(n-\frac{3}{2}\right) 
        \,\left\{\beta\int \,P(x)\,d^3\xx -N \right\}.
\label{Appen_A: tsallis_entropy}
\end{eqnarray}
Thus, the substitution of integral formula (\ref{Appen_A: integral}) 
immediately leads to 
\begin{eqnarray}
S_q  = \left(n-\frac{3}{2}\right)\left[\frac{1}{n-5}\,
  \left\{8\pi r_e^3 P_e-(n+1)\frac{MP_e}{\rho_e}
        +\frac{GM^2}{r_e}\right\}\beta +N\right],
\nonumber
\end{eqnarray}
which can be expressed in terms of the homology invariants: 
\begin{eqnarray}
 \label{Appen_A: entropy_uv}
 S_q = \left(n-\frac{3}{2}\right)
 \left[\,\frac{1}{n-5}\,\,\frac{\beta GM^2}{r_e}\,
   \left\{2\frac{u_e}{v_e}-(n+1)\frac{1}{v_e}+1 \right\} +N\right].
\end{eqnarray}
%
%
%
\subsection*{A.3~ Radius-mass-temperature relation }
%
%
%
The mass-radius-temperature relation (\ref{eq: eta}) is 
derived from the equilibrium stellar polytropic configuration. 
Using (\ref{eq: hydro_1}), we first write down 
the condition of hydrostatic equilibrium at the boundary $r_e$: 
\begin{eqnarray}
  \frac{GM}{r_{e}^{2}}=-\frac{1}{\rho_e}\,\left(\frac{dP}{dr}\right)_e.
\nonumber
\end{eqnarray}
The right-hand-side of this  equation is rewritten with 
dimensionless quantities in (\ref{eq: dimensionless}): 
\begin{equation}
  \label{Appen_A: hydro2}
  \frac{GM}{r_{e}^{2}}=-(n+1)K_{n}\,\rho_c^{1/n}\,\,
  \left(\frac{\xi_e}{r_e}\right)\,\,\theta'_e.
\end{equation}
We wish to express the above equation only in terms of the variables 
at the edge. To do this, we eliminate the residual dependences, 
$\rho_c$ and $K_n$ from (\ref{Appen_A: hydro2}). The definition 
(\ref{eq: dimensionless}) leads to 
\begin{eqnarray}
\frac{\xi_e}{r_e}=\left\{ \frac{4\pi G \rho_c^2}{(n+1)P_c}\right\}^{1/2}
=\left\{ \frac{4\pi G}{(n+1)K_n} \right\}^{1/2}\,\,\rho_c^{(n-1)/(2n)}, 
\nonumber
\end{eqnarray}
which can be rewritten with 
\begin{eqnarray}
\rho_c^{1/n} = \left\{\frac{4\pi G}{(n+1)K_n}\right\}^{1/(n-1)}\,\,
        \left(\frac{\xi_e}{r_e}\right)^{2/(n-1)}.  
\nonumber
\end{eqnarray}
Substituting the above relation into (\ref{Appen_A: hydro2}), the 
$\rho_c$-dependence is first eliminated and we obtain
\begin{equation}
\frac{G^{n/(n-1)}M}{r_e^{(n-3)/(n-1)}}=
        -\left[\frac{\left\{ (n+1)K_n \right\}^n}{4\pi}\right]^{1/(n-1)}\,\,
        \xi_e^{(n+1)/(n-1)}\,\theta'_e.
\label{Appen_A: gmr_relation}
\end{equation}
As for $K_n$-dependence, the definition (\ref{eq: def_Kn}) together with 
(\ref{eq: A_Phi0_constants}) yields 
\begin{eqnarray}
(n+1)\,K_n = 
        \left\{ 4\sqrt{2}\pi\,\,
\frac{B(3/2,n-1/2)}{(n-1)^{n-3/2}}\,\,
\frac{M}{h^3}\right\}^{-1/n}(m_0\beta)^{-(n-3/2)/n}.
\label{Appen_A: Kn-beta}
\end{eqnarray}
Hence, substituting the above expression into (\ref{Appen_A: gmr_relation}),  
the relation between mass $M$, radius $r_e$ and Lagrange multiplier $\beta$ 
can be finally obtained. In terms of the homology invariants, it follows that  
\begin{equation}
\left\{\frac{(GM)^{n}(m_0\beta)^{n-3/2}}{r_e^{n-3}h^{3}}\right\}^{1/(n-1)}
\,\,=\,\,\alpha_n\,\left(u_e\,\,v_e^n\right)^{1/(n-1)}, 
\end{equation}
where the constant $\alpha_n$ is given by
\begin{eqnarray}
\alpha_n \,\,\equiv\,\, \left\{\frac{(n-1/2)^{n-3/2}}{16\sqrt{2}\pi^2\,
(n+1)^n\,B(3/2,n-1/2)}\right\}^{1/(n-1)},  
\nonumber
\end{eqnarray}
which asymptotically approaches unity in the limit $n\to\infty$. 
%
%
%
%
%
%
%
%
\section*{Appendix B: Thermodynamic temperature of classical gas 
  model from the Carnot cycle}
\label{appen_B}
%
%
%
%
%
%
%
%
In a standard framework of thermodynamics, the temperature is defined 
by means of an efficiency of the Carnot cycle. Here we apply the standard 
procedure to seek the physical temperature $T_{\rm phys}$ for so-called 
polytropic system of which distribution function is given by the extremization
of the Tsallis entropy (see eqs.(\ref{eq: S_q})(\ref{eq: 1st_variation})).
For simplicity, we discuss a case of the free classical gas 
without gravity, which corresponds to the $G \rightarrow 0$ limit of 
the {\it stellar polytropic system}.

From the $G \rightarrow 0$ limit of the formula (\ref{Appen_A: energy}), 
free polytropic system of the volume $V$ with {\it homogeneous} 
pressure $P$ and density $\rho$ has an (internal) energy: 
\begin{equation}
E = K = \frac{3}{2}\, P V  = \frac{3}{2}\,\frac{MP}{\rho}. 
\label{Appn_B: energy}
\end{equation}
Here we drop the subscript ${_e}$ for the pressure and density, since both are 
constant within the system in absence of gravity. 
And equation of state (\ref{eq: Eq. of state}) becomes  
\begin{equation}
P = K_n ~\rho^{1+1/n}~ = K_n \, \biggl(\frac{M}{V}\biggr)^{1+1/n}.
\label{Appen_B: eq of state}
\end{equation}
From equations (\ref{eq: A_Phi0_constants}) and (\ref{eq: def_Kn}), 
the constant $K_n$ is related to the Lagrange multiplier $\beta$ as 
\begin{equation}
K_n \propto \beta^{-(n-3/2)/n},
\label{Appen_B: rel Kn to beta}
\end{equation}
so that this constant can be used as a parameter which characterizes 
the temperature of the system. However, it is not sure whether 
$K_n$ itself has a role of the physical temperature, which should be 
determined through the efficiency of the Carnot cycle.

The internal energy (\ref{Appn_B: energy}) and the equation of state 
(\ref{Appen_B: eq of state}) give the thermodynamic first law: 
\begin{eqnarray}
d'Q &=& \,dE + \, P \, dV \nonumber \\
    &=& \,M^{1+1/n} \left\{\,
\frac{3}{2}\, \frac{dK_n}{V^{1/n}} \, + \,
\left(\frac{n-3/2}{n}\right) K_n\,\,\frac{dV}{V^{1+1/n}}
\right\}\, , 
\label{Appen_B: dQ}
\end{eqnarray}
from which adiabatic changes $d'Q = 0 $ is expressed as
\begin{eqnarray}
K_n\, V^{(2/3 - 1/n)} = {\rm constant}, ~~~~P\, V^{5/3} = {\rm constant'}.
\label{Appen_B: Adiabatic}
\end{eqnarray}
Note that adiabatic lines in a $P$-$V$ plane become 
steeper than isothermal ones when $n > 3/2$.

\begin{figure}
  \begin{center}
    \includegraphics*[width=10cm]{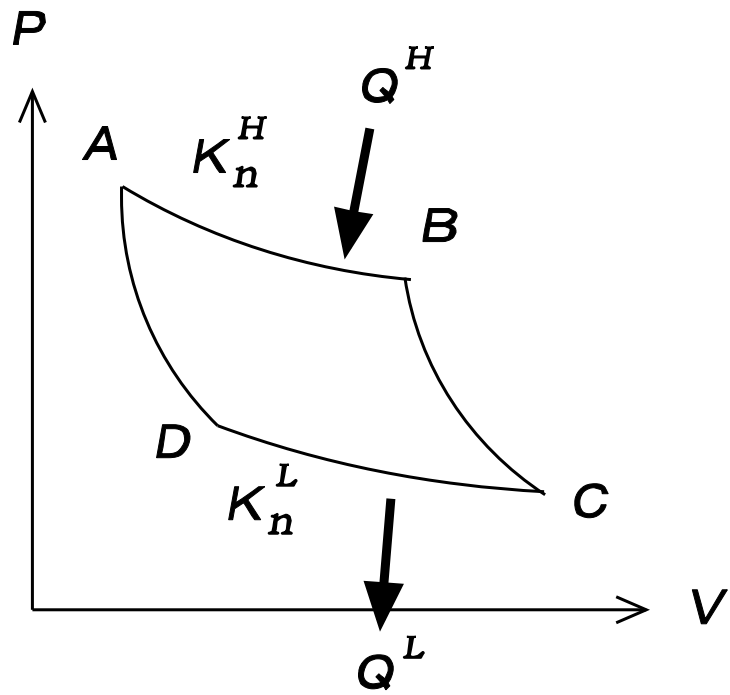}
  \end{center}
    \caption{A schematic description of Carnot cycle. 
      \label{fig: Carnot}}
\end{figure}

Now, let us consider the Carnot cycle shown in Figure \ref{fig: Carnot}. 
As usual, quasi-static changes  
B $\rightarrow$ C and D $\rightarrow$ A are adiabatic. As for the process 
A $\rightarrow$ B, the system is in a thermal contact with a heat bath which 
has a higher temperature $K_n^H$. Similarly, during the change 
C $\rightarrow$ D, the system lies in a thermal equilibrium with another
heat bath that has a lower temperature $K_n^L$.  
The system absorbs amount of heat $Q^H$ from the higher temperature 
bath and disposes $Q^L$ to the lower one during the isothermal processes 
A$\rightarrow$ B and C$\rightarrow$ D, respectively. They 
are easily evaluated from (\ref{Appen_B: dQ}):   
\begin{equation}
\begin{array}{ll}
Q^H &= (n-\frac{3}{2})\,M^{1+1/n}\,\, K_n^H \,\,
\Bigl(V_A^{-1/n}- V_B^{-1/n} \Bigl) ,  \\
Q^L &= (n-\frac{3}{2})\,M^{1+1/n}\,\, K_n^L \,\,
\Bigl(V_D^{-1/n}- V_C^{-1/n} \Bigl) . 
\end{array}
\label{Appen_B: heat}
\end{equation} 
On the other hand, a relation between the parameters of the cycle 
can be obtained from the equation of state (\ref{Appen_B: eq of state}) 
and the adiabatic changes (\ref{Appen_B: Adiabatic}): 
\begin{eqnarray}
\Biggl(\frac{K_n^H}{K_n^L}\Biggl)^{\gamma} = \frac{V_C}{V_B} = 
\frac{V_D}{V_A}~;~~~\gamma = \frac{3}{2}\,\,\frac{n}{n-3/2}.
\label{Appen_B: relations}
\end{eqnarray}
Thus, equations (\ref{Appen_B: heat}) and (\ref{Appen_B: relations}) 
lead to the following efficiency of the Carnot cycle:  
\begin{equation}
\eta \equiv 1 - \frac{Q^L}{Q^H} =\,\, 
1 \,\,- \,\,\Biggr(\frac{K_n^L}{K_n^H}\Biggl)^{n/(n-3/2)} 
=\,\, 1 \,\,- \,\,\frac{\beta^H}{\beta^L},
\end{equation}
where we used the relation (\ref{Appen_B: rel Kn to beta}) in the last line. 
This clearly shows that the inverse of the Lagrange multiplier $\beta$ 
has a role of the physical temperature. 
\clearpage
%
%
%
%
%
%

%
%
%
%
%
%
%
%
%
%
%
%
%
%
%
%
%
%
%
%
%
%
%
%
%
%
%
\clearpage
%
%
%
%
%
%
\begin{table}
\caption{Critical values of the radius-mass-temperature relation, 
$\eta_{\rm crit}$ and the density contrast between center and edge, 
$D_{\rm crit}=(\rho_c/\rho_e)_{\rm crit}$ in the case of a system in contact 
with a heat bath for given polytrope index $n$ or $q$.}
\label{tab: ErGM}

\vspace*{0.5cm}

  \begin{center}
\begin{tabular}{|ccccc|} 
\hline
 \makebox[1.0cm]{n}  & \makebox[1.0cm]{q} &\makebox[0.1cm]{} &
 \makebox[2.5cm]{$\eta_{\rm crit}$} 
& \makebox[2.5cm]{$D_{\rm crit}$}\\ 
\hline\hline
3 & $\frac{5}{3}$ && ----- & ----- \\
4 & $\frac{7}{5}$ && 0.9421 & 153.5 \\
5 & $\frac{9}{7}$ && 1.193 & 88.15 \\
6 & 1.22 &&  1.379 & 68.38 \\
7 & 1.18 &&  1.520 & 58.86 \\
8 & 1.15 &&  1.631 & 53.28 \\
9 & 1.13 &&  1.720 & 49.62 \\
10 & 1.12 &&  1.793 & 47.04 \\
30 & 1.04 &&  2.263 & 35.89 \\
50 & 1.02 &&  2.363 & 34.28 \\
100 & 1.01 &&  2.440 & 33.17 \\
$\infty$ & 1 &&  2.518 & 32.13
\\ \hline
\end{tabular}
  \end{center}
\end{table}
%
%
%
%
%
%
%
\clearpage
%
%
%
%
%
%

\vspace*{0.5cm}

\vspace*{0.8cm}

\vspace*{0.8cm}

%
%
%
%
%
%
\end{document}